# Metabolomic and flux-balance analysis of age-related decline of hypoxia tolerance in *Drosophila* muscle tissue

Laurence Coquin[*1], Jacob D. Feala[*2], Andrew D. McCulloch[2], Giovanni Paternostro[1,2]

* contributed equally.
[1]Burnham Institute, 10901 North Torrey Pines Road, La Jolla, CA 92037
[2]Department of Bioengineering, University of California, San Diego, 9500 Gilman Drive, 0412, La Jolla, CA 92093-0412

**Corresponding author:**
Giovanni Paternostro
The Burnham Institute
10901 North Torrey Pines Road
La Jolla, CA 92037
Tel. (858) 713 6294
Fax (858) 713 6281
e-mail: giovanni@burnham.org

## Abstract

The fruit fly *D. melanogaster* is increasingly used as a model organism for studying acute hypoxia tolerance and for studying aging, but the interactions between these two factors are not well known. Here we show that hypoxia tolerance degrades with age in post-hypoxic recovery of whole-body movement, heart rate and ATP content. We previously used $^1$H NMR metabolomics and a constraint-based model of ATP-generating metabolism to discover the end products of hypoxic metabolism in flies and generate hypotheses for the biological mechanisms. We expand the reactions in the model using tissue- and age-specific microarray data from the literature, and then examine metabolomic profiles of thoraxes after 4 hours at 0.5% $O_2$ and after 5 minutes of recovery in 40- versus 3-day-old flies. Model simulations were constrained to fluxes calculated from these data. Simulations suggest that the decreased ATP production during reoxygenation seen in aging flies can be attributed to reduced recovery of mitochondrial respiration pathways and concomitant over-dependence on the acetate production pathway as an energy source.

**Subject Categories:** Metabolic and regulatory networks, Cellular Metabolism.

# Introduction

Aging is universal to eukaryotic organisms, and at the cellular level its effects are global, reaching virtually all cellular processes. Normal function deteriorates with age, but more dangerous is the loss of ability to respond to external stresses, contributing to a higher risk of death from external causes (Rose, 1991). Of particular interest is the age-related decline of cellular hypoxia tolerance, since hypoxic damage to heart and brain tissue is the source of pathology in heart attacks and strokes. In the heart, both the incidence and mortality of ischemic events worsen with age (Bonow et al., 1996). Currently there is a need for preventative measures to improve tolerance to ischemia-reperfusion injury in high-risk patients. The fruitfly *Drosophila* presents a possible source of new discoveries, since it shares fundamental oxygen regulation pathways with humans but is highly tolerant to hypoxia stimuli (and accompanying reoxygenation as well) (Krishnan et al., 1997). *Drosophila* is also a common model for aging research, and relationships have begun to be explored between aging and chronic hypoxia tolerance (Vigne & Frelin, 2007), and between aging and oxidative stress (Zou *et al.*, 2000). However, the interaction between aging and acute hypoxia tolerance in flies has not yet been investigated.

The exact mechanism whereby reversible hypoxic tissue damage finally evolves into irreversible damage is still controversial (Opie, 1998a)**,** but is likely to involve both necrotic and apoptotic mechanisms, stemming from metabolic stresses introduced during reperfusion as well as the ischemic event itself. Reduced $O_2$ causes reduction in oxidative metabolism and increased dependence on glycolysis. Under normal conditions of mild hypoxia and steady ATP demand, such as in highly active muscle tissue, consumption of protons by oxidative pathways cannot keep up with protons produced by ATP hydrolysis and a reversible state of acidosis results (Robergs et al., 2004). Contractile machinery and metabolic enzymes are negatively regulated by acidosis (Opie, 1998a). In severe hypoxia, ion pumps are inhibited by depletion of ATP. Ion pump inhibition causes decreased uptake of calcium by the sarcoplasmic reticulum and reduced extrusion from the cell (Steenbergen et al., 1987), and this calcium accumulation can damage mitochondria. Upon reperfusion, the cell experiences sudden oxygen influxes that its inactive oxidative pathways and damaged mitochondria cannot immediately metabolize, resulting in the creation of reactive oxygen species (Ambrosio et al., 1987). It has been suggested that one way that hypoxia tolerant organisms prevent these dangerous imbalances by rapid and global regulation of metabolism (Hochachka, 1980; Hochachka, 2003).

All animals have complex, multiscale systems for regulating oxygen homeostasis (Hochachka & Somero, 2002). At the cellular level, hypoxia resistance mechanisms most likely evolved very early and appear to be highly conserved among species (O'Farrell, 2001). Supporting this hypothesis, several fly genes have been discovered that are similar in sequence and function to human genes for regulation of metabolism, signaling, and transcription during hypoxia (Lavista-Llanos et al., 2002; Pan & Hardie, 2002; Wingrove & O'Farrell, 1999). Although the hypoxia response in flies and humans seems to share similarities at the level of individual genes, stark contrasts exist at the phenotype level. Flies have a remarkable tolerance to hypoxia that is the subject of an increasing amount of investigation. In contrast with humans, who can only survive a few minutes without oxygen, flies can fully recover from up to 4 hours in complete anoxia (Krishnan et al., 1997). Genetic determinants of fly hypoxia tolerance have been discovered by genetic screens (Haddad et al., 1997a), and one enzyme unique to *Drosophila* increased hypoxia tolerance when transferred to human cells (Chen et al., 2003). Flies' innate hypoxia tolerance can be further improved by directed evolution, and gene deletions mimicking these evolutionary changes in gene expression improve tolerance in wild-type flies (Zhou et al., 2007).

The fruitfly is also one of the principal model organisms used for studying the genetics of aging, for a number of reasons. Flies develop to adulthood quickly, have a short life span, and share a number of characteristics of functional senescence with humans (Grotewiel et al., 2005). Little is known about how aging degrades hypoxia defences. However, aging is well known to profoundly affect metabolism, and metabolism plays a central role in genetic interventions on the aging process. In fact, many of the best characterized genes that can accelerate or retard aging in model organisms act on the insulin pathway and on mitochondria and are studied in flies (Giannakou & Partridge, 2007). Heart and muscle tissue are good indicators of functional senescence since they have a single functional purpose (contraction) which is highly dependent on metabolic regulation, and they have easily quantifiable phenotypes (heart rate and physical activity). Another advantage of using fruitflies to study muscle tissue biochemistry is that the fly thorax is composed primarily of flight muscle by mass, which allows for easy dissection of relatively large numbers of flies with fairly high specificity for muscle tissue. (However, it must be noted that an unavoidable limitation of using whole thoraxes is that there will be some contamination with hemolymph and other tissue types.)

The physiological effects and responses to extreme oxygen conditions can manifest on many biological levels. Because metabolites are downstream of gene transcripts and proteins, changes in metabolite levels can provide an indication of the overall integrated response of an organism. To obtain a better understanding of the system-wide effects of hypoxia on fly muscle tissue, we use NMR spectroscopy to simultaneously measure many metabolites present in the tissue and their changes in response to hypoxic stress, as described previously (Feala et al., 2007). This approach, termed metabolomics, is complementary to genomics and proteomics in studying the complex biological system response to chemical, physical, and genetic factors (Goodacre et al., 2004; Griffin, 2003; Griffin & Bollard, 2004; Nicholson et al., 1999). The simultaneous measurement of a large number of metabolites, in combination with the use of a constraint-based computational model, allows us to quantify global changes in metabolite fluxes.

The present study investigates how aging affects the metabolic response to hypoxia in *Drosophila melanogaster*. First, we exposed young and old flies to severe hypoxia and compared the age-related degradation in physiological recovery at the levels of the organism (whole-body activity), organ (heart rate), and cell (ATP content). Then, for both age groups, we gathered metabolite profiles during hypoxia and recovery and compared these to an untreated control. Metabolite fluxes were calculated for hypoxia and recovery and integrated into the model, then simulations of network function were inspected for differences in key fluxes such as ATP, $H^+$, and glucose. This analysis generated hypotheses for mechanisms of the loss of hypoxia tolerance with age, and these hypotheses were checked for consistency against existing transcription profiles of young and old flies (Girardot et al., 2006). The results show the utility of NMR metabolomic profiling for characterization of the instantaneous physiological condition, enabling direct visualization of the perturbation of, and return to, homeostasis.

## Results

### *Post-hypoxic recovery of physiological function*

The advantages of using Drosophila to study mechanisms of aging are further enhanced by the many similarities in age-related degradation of function between flies and humans. For example, we previously found that flies experience a decline in maximum heart rate with age that is similar to humans (Paternostro et al., 2001). Grotewiel et. al. (2005) review other age-related declines in flies, including motor activity, stress response (including oxidative stress), and ATP production. We examined the senescence of the physiological response to hypoxia in three different experiments on young (3-day-old or "3-day") and old (40-day-old or "40-day") flies.

Flies respond to acute hypoxic stress by falling into a motionless, prostrate stupor, from which they can fully recover after several minutes (Haddad et al., 1997b). Aging significantly delayed the recovery of whole-body activity after extreme hypoxic stress (4 hours at 0.5% O2, N = 16 males each group) according to Student's t-Test. Figure 1A depicts the cumulative recovery to standing position for each group, with Kaplan-Meier estimates of 95% confidence intervals. Young flies began to return to standing position after an interval of 32 minutes post-hypoxia, with approximately 2/3 arousing within the first 2 hours. Old flies remained motionless for the first 4 hours post-hypoxia, with 2/3 arousing within 8 hours. After 24 hours, the percent of fully recovered flies was equivalent between the two age groups.

Recovery of heart activity after the same treatment followed a similar trend, though on a different timescale (shown in Figure 1B, N = 11 males for each group). The fly has a tube-like heart that normally contracts at 6 to 8 beats per second (bps). At the end of the hypoxic period, the hearts of both young and old flies were completely stopped. In young flies, the heart began slowly beating within the first minute, increasing quickly to approximately 3 bps in the third minute, then maintaining a range between 2 and 4 bps for the remainder of the 20 minute measurement duration. In contrast, older hearts remained mostly motionless for the first 6-7 minutes, and then steadily recovered over the remaining interval (see Supplementary Materials). Heart rates for both groups eventually recovered to baseline levels, but this happened over a longer timescale than presented here (the measurement duration was limited for technical reasons). Inverse beat-to-beat intervals binned over the first 5 minutes of measurement and normalized to baseline were significantly different ($p < 0.05$) by Student's t-test (Figure 1A).

Similarly, we measured ATP concentrations in flies at baseline, at the end of a 4-hour hypoxia stimulus and after a 5-minute recovery period – reflecting the time period seen in the heart recovery data (Figure 2). Under normal oxygen, ATP concentrations are 1.8-fold higher in 3-day than 40-day flies (P=0.006). However, at end of the hypoxia treatment, ATP levels are very low and equivalent in the two age groups. When the flies are allowed to recover, ATP concentration is 6.6-fold higher in 3d flies than 40d ones (P<0.001).

## *Metabolite Assays*

### Glycogen, glucose and trehalose

Glycogen, free cellular glucose, and trehalose are the major sources of carbohydrate fuel in flight muscle in many Diptera such as bees and blowflies (Childress et al., 1970; Suarez et al., 2005), and are likely to be for *Drosophila melanogaster* as well. The large deposits of glycogen in flight muscle of flies, the depletion of these reserves after prolonged flights, and the rapid catabolism of the polysaccharide by flight muscle *in vitro*, indicate that glycogen provides a major vehicle for storage of sources of potential energy which can be mobilized to meet the metabolic requirements of active muscle (Sacktor & Wormser-Shavit, 1966). The disaccharide trehalose can also support flight activity; it was identified as the principal blood sugar in many species of insects, was found in muscle, was found to be reduced in concentration within these loci after flight, and was metabolized in vitro by flight muscle.(Sacktor & Wormser-Shavit, 1966)

Glycogen and trehalose concentrations are difficult to quantify by our NMR assay. Trehalose, although visible in the spectra, binds proteins with high affinity and thus a highly variable proportion is filtered from the supernatant along with the soluble proteins. Glycogen can also be seen in the spectra, but cannot be quantified due to the variable lengths of each polymer chain. Therefore, these important substrates were measured biochemically, following enzymatic assays developed by Parrou (Parrou & Francois, 1997).

As for ATP, we measured glycogen concentrations in flies at baseline, at the end of a 4-hour hypoxia stimulus, and after a 5-minute recovery period (Figure 3). Glycogen was found to be the major source of fuel used by young and old flies to produce glucose under hypoxic conditions, with concentrations decreasing greatly as the substrate was consumed over the hypoxia duration. In both age groups, hypoxic trehalose levels were not statistically different from the ones measured under normoxia, and further there were no significant differences across age groups for the two treatment conditions. Old flies showed consumption of glycogen and trehalose during the recovery period ($p = 0.004$ for glycogen and $p = 0.005$ for trehalose.)

### $^1$H NMR metabolomics

NMR spectra (as shown in Figure 1-S of the Supplementary Material) were used to quantify free metabolite concentrations in samples of 20 fly thoraxes, homogenized, filtered of protein, and buffered in 500μL of $D_2O$ (see Materials and Methods). Of 37 metabolites identified, 26 had at least one measurement higher than our measurement threshold of 10µM (in the $D_2O$ homogenate) (see Table I-S and Table II-S for summaries, and the spreadsheet "aging_hypoxia_NMR.xls" for raw data, in the Supplementary Material). Free glucose was measured by both NMR and biochemical assays, allowing us to check for consistency in the data. Glucose concentrations had similar qualitative behavior in both datasets, increasing in young flies during hypoxia and then returning toward baseline levels during recovery, while in old flies remaining steady during hypoxia but decreasing during recovery.

Metabolomics experiments generate multivariate data, which complicate statistical analysis by typically having a larger number of variables than experimental samples. Principal component analysis (PCA) is a vector transformation that can reduce this high dimensionality by projecting the data "cloud" (each point in the cloud representing a data sample) onto new axes in the multivariate space. The new axes are an orthogonal set of basis vectors that are a weighted composite of the variables (in this case, metabolites).

We applied PCA to the metabolomic profiles of young and old flies. When plotted on their principal components (Figure 4A), young and old flies had very similar profiles in control and hypoxic conditions, with large shifts on PC 2 during 4 hours hypoxia and smaller movements in the direction of PC 1. After 5 minutes in room air, both groups returned toward controls along the direction of PC 2, whereas young flies continued to drift slightly along PC 1. Older flies had a pronounced movement along PC1 during recovery, corresponding to the large increase in acetate seen in NMR data.

Decomposition of the data by PCA captured nearly 80% of the variability of the concentration data with the first principal component (PC), with the second PC contributing another 15% to the total variability. Acetate production dominated PC 1, while alanine and lactate production were responsible for most of the changes on PC 2. Oxalacetate, glutamate, glucose, and glutamine had minor contributions to the two PCs (Figure 4B). Although similar conclusions can be drawn from direct inspection of the NMR profiles, PCA confirmed that these changes were the main sources of variability across the 6 datasets. In addition, the data show that acetate production is orthogonal to alanine and lactate, and may therefore be attributed to separate regulatory mechanisms.

One-way analysis of variance evaluated the effect of hypoxia treatment and recovery, independent of the age groups. Out of the 26 metabolites with concentrations above the 10µM threshold, only 10 were affected by the treatment ($p < 0.047$ with Bonferroni correction). As we reported previously, lactate, alanine and acetate are the major end products of hypoxic metabolism in Drosophila, and again were the only metabolites with large increases over the hypoxic duration in both young and old flies. On reoxygenation, fluxes reverse quickly and increase several hundred-fold, and we observed that most of the metabolites returned to control levels within the 5-minute measurement duration. However, as suggested by the PCA analysis, the acetate level continued to increase during recovery in both young and old flies.

We also employed the Student t-test to individually identify metabolites that significantly vary between old and young flies for each treatment (see Table III-S: *p*-values for t-tests in the Supplementary Material). We noticed that during normoxia and hypoxia few metabolites vary significantly between age groups, even before Bonferroni correction for the number of metabolites. The highest confidence differences between the ages appear during the recovery period. After 5 minutes recovery under normal oxygen conditions, the lactate level returns to its normal value in young flies whereas in old flies it is still increased by 80% compared to control ($p = 0.005$). Both young and old flies continue to produce acetate during recovery, though old flies produce much more (young: +534% compared to control; old: +1800% compared to control, $p = 0.005$). It is also interesting to note that during recovery in old flies, oxalacetate concentration increased by 230% compared with control.

During hypoxia, the production of alanine is not matched by consumption of proline, as it is during aerobic exercise in insect flight muscle. In fact, amino groups do not balance in hypoxia or recovery for either age group, therefore protein degradation and formation may be a factor, respectively, under these conditions. Production or consumption of free amino groups was calculated during flux-balance analysis, described below.

*Metabolic reconstruction*

In order to further refine our existing genome-scale metabolic reconstruction (Feala et al., 2007) for muscle tissue, we created a filtered gene list based on global gene expression measured in 3-day-old thoraxes by Girardot (2006). Since thorax tissue is composed mostly of flight muscle, highly expressed enzyme genes in this dataset could be added with confidence to our metabolic network. These data were measured on Affymetrix microarrays, which provide absolute measurements of mRNA levels.

The histograms in Figure 5 display the distribution of the microarray data after filtering and integration with the KEGG Pathway Database. The distribution of thorax genes linked to at least one KEGG enzyme (Figure 5B) has an interesting bimodal distribution which is much less prominent in the histogram of all genes (Figure 5A). This "long tail" roughly corresponds to the threshold of expression (500) for inclusion into the model, which was determined empirically from a literature and database search of samples of genes at all levels of expression. The distribution of mean expression level for all KEGG pathways is shown in Figure 5C. The right tail of this distribution also seems to correspond to pathways known to be active in flight muscle tissue, as exemplified by the pathways labeled in the figure. Table 1 notes new pathways that were included in the model based on mean expression level. In all, 49 new genes and 38 new reactions were added to the model from the previous version, resulting in totals of 211 genes and 196 reactions. In addition, the new version contains many minor improvements to existing reactions, such as cellular compartment assignments and gene-protein-reaction associations, as well as the removal of enzymes and pathways with low expression levels. A complete map of the network is shown in Figure 2-S of the Supplementary Materials, along with a list of reactions in "model_reactions.xls".

*Flux-balance analysis*

Though a few *Drosophila* enzymes have been extensively studied, most of the reaction kinetics in fly central metabolism are unknown. To capture as wide a scope of pathways as possible, while avoiding the necessity of kinetic parameters, we instead applied flux-balance analysis to our network reconstruction. The major assumption in flux-balance analysis is that the system is at steady state, therefore intra-system metabolite concentrations do not change.

In addition to the steady state requirement, our analysis assumes that accumulation and depletion of metabolites are linear and unidirectional, i.e. fluxes are constant over the measurement period and there is no consumption of accumulating metabolites or synthesis of depleted metabolites. The fact that the heart stops beating during hypoxia supports the assumption that the cells use only the carbohydrate stores that we measure, and nutrients are not supplied by the fat body via recirculated hemolymph.

The model was constrained using data from eleven metabolites which were chosen based on the magnitude of the changes during hypoxia and recovery and the existence of each metabolite within the network. To increase accuracy of the simulations, absolute concentrations were estimated from the NMR spectra using correction factors as described in the Methods. For each metabolite selected, we approximated a flux by dividing the concentration differences by the experimental time period (4 hours for hypoxia, 5 minutes for recovery). Figure 6 (top row) shows estimated fluxes of these metabolites for each condition, derived from concentration data in the NMR and biochemical assays. Exchange reactions were added to the model and for each simulation their fluxes into or out of the system were constrained to approximate rates of accumulation or depletion.

Using the mean and standard error calculated for each flux, Gaussian distributions were constructed and sampled to create 10,000 sets of the eleven metabolite fluxes, which were then applied to the model to analyze the sensitivity of simulations to variation in the NMR data. All populations of fluxes for each reaction had approximately normal distributions. Figure 6 (bottom row) shows intrasystem fluxes calculated in flux-balance simulations. The variance of calculated fluxes is probably conservative in that each measured flux is treated as an independent random variable during sampling even though in the biological system there are likely many correlations among metabolites that are disregarded here. Intrasystem fluxes in the model follow the same pattern as the NMR data, with wide variances during hypoxia but much tighter distributions during recovery.

Except for glutamate degradation and the resultant production of ammonium, hypoxic fluxes were not significantly different between old and young flies (Figure 6 – left). The opposite signs of measured oxaloacetate fluxes drive a small set of anaplerotic reactions, involving glutamate, in different directions; however, the fluxes are small in both age groups. Young flies consume more free ammonium, driven by surplus production of alanine shown in the NMR data. However, some protein degradation, which was not included in this model, may partially account for the appearance of $NH_4$ not provided by free amino acids. Glycolysis and TCA cycle pathways are used at the same rate in young and old, although production of lactate is slightly higher in the old flies. Proton production and accumulation of anaerobic end products is high for both age groups. Surprisingly, calculations of ATP production during the hypoxic period were the same in old and young flies.

During recovery, intrasystem fluxes in the model show more drastic differences between the ages, especially in the recovery of oxidative pathways. Old flies show much higher acetate production during recovery, which causes the model to calculate slightly higher glycolytic fluxes and much lower TCA cycle fluxes due to the difference in acetyl-CoA conversion to acetate *versus* oxidation via the TCA cycle. As the flux map of Figure 7 shows, reduced activity of the TCA cycle diverts acetyl-CoA through the acetate pathway. Differences in ATP production during recovery are significant; however, the additional ATP from the cleavage of CoA makes up for some of the resulting loss of efficiency in carbon utilization. Due to the slow recovery of oxidative metabolism, proton fluxes are negligible in old flies, a marked difference from the proton consumption that occurs during recovery in 3-day flies. Table II shows key fluxes and ratios from the simulations.

## Discussion

Aging causes deterioration of a variety of physiological functions at the organism level, which originate from global degradation and dysregulation at the cellular and molecular scale. The nature of aging as a genome-wide, multiscale phenomenon invites a systems-level approach to understanding the cellular mechanisms by which phenotypes degrade. In this study we have focused on hypoxia tolerance in Drosophila, which is notably higher than that of mammals but degrades similarly with age, as we show using three different physiological measurements at three different scales. Though our physiological phenotype measurements were not significantly different between young and old flies during hypoxia, upon reoxygenation the age groups had major differences. At the organism level, old flies have slower whole-body recovery of activity following hypoxic stress; at the organ level, post-hypoxic heart rate takes longer to recover in older flies; and at the cellular level, ATP levels are slower to recover after hypoxia in old flies.

We used metabolomic and computational analysis of muscle metabolism to search for clues as to the molecular basis for this delayed recovery. Global metabolite profiles were gathered during baseline, after a long hypoxic period, and then after a short recovery as in the physiological measurements, and the data were entered into our microarray-supported stoichiometric model of Drosophila central metabolism. Flux-balance analysis within this model, driven by flux estimates from the metabolic profiles, reiterated that the major differences in metabolic function between young and old flies occur on reoxygenation rather than during hypoxic stress. Further, the model also predicts that differences in recovery of mitochondrial respiration, and the resulting effects on proton production and glucose utilization in old flies, may contribute to the differences in physiological recovery.

Aging is often described as a generalized deterioration of function but our results show that not all metabolic pathways are equally affected. The impairment in metabolic recovery after hypoxia seems to be mainly in pathways downstream of pyruvate (Krebs cycle and respiration) rather than in the anaerobic portion of glycolysis. In our model, the decline in respiratory function seems to account for most of the difference in hypoxia tolerance with age, which is widely supported by evidence that mitochondrial function plays a major role in the overall functional decline seen with age in *Drosophila* (Dubessay et al., 2007; Ferguson et al., 2005) similar to its role mammals (Kujoth et al., 2005; Martin, 2001; Martin et al., 2003; Trifunovic et al., 2004). Activity of electron transport chain enzyme complexes I, III, and IV decrease with age in flies although the expression of certain protein subunits of these complexes remains unchanged (Dubessay et al., 2007; Ferguson et al., 2005). Microarray (Girardot et al., 2006; Zou et al., 2000) and Northern blot (Dubessay et al., 2007) measurements of RNA expression suggest that transcript levels of TCA cycle and respiratory enzymes are highly downregulated with age. Glucose and glycogen consumption is necessarily higher, since all other catabolic pathways are less efficient in terms of ATP production. The decreased ADP/O ratios seen in mitochondrial assays of old flies (Dubessay et al., 2007) were not included in our model simulations, but would most likely only accentuate these results.

The specific cause of this decline in respiratory metabolism is still under investigation. Multiple intermittent reperfusions during anoxia causes injury in young flies, marked by lower rates of respiration on reoxygenation (Lighton & Schilman, 2007), which supports the hypothesis that fly mitochondria can be damaged by reactive oxygen species (ROS) created by oxygen reperfusion. In older flies, reduced respiration has been attributed to both a lifetime of accumulated damage from (Dubessay et al., 2007; Ferguson et al., 2005), and a chronically active response to (Zou et al., 2000), the generation of ROS in the mitochondria.

Another possible contributor to the physiological response to hypoxia in heart and muscle tissue is acidosis. The production of lactate, alanine, or acetate end products from pyruvate partially uncouples glycolysis from oxidative metabolism, causing an imbalance in proton production (by ATP hydrolysis) and consumption (by ATP synthase) (Robergs et al., 2004). Since acidosis negatively regulates contractility both at the sarcoplasmic reticulum and actin-myosin interaction (Opie, 1998b), heart rate and muscular activity would be expected to recover faster in the system that is quicker to reverse proton accumulation. Young and old flies produce an equivalent amount of protons during hypoxia, suggesting that acidosis is unavoidable even in hypoxia-tolerant organisms. However, after 5 minutes of reoxygenation, protons in 3-day flies are being consumed at a high rate by ATP synthase, while the model calculates nearly zero proton flux in 40-day flies. This, in combination with the lower rate of ATP production in old flies according to the model, can help to explain our observations of age-related differences in recovery of physiological functions. Our ATP assay also confirms slower restoration of ATP levels in older flies.

In anaerobic pathways, the major difference between old and young flies is in the production of acetate. Out of the three end products lactate, alanine, and acetate, acetate is the only compound still being produced during recovery, and the reason might be that the additional ATP and NADH per glucose created by this pathway result in a better ATP/H$^+$ ratio than that of the other two pathways. Therefore, acetate production, in both young and old flies, may represent the most efficient utilization of any surplus pyruvate that exceeds the oxidative capacity of the recovering mitochondria at 5 minutes post-reperfusion. The model, which solves for the optimal flux distribution for ATP production, supports this hypothesis by converting all pyruvate to acetate when oxygen is restricted, and given a choice of anaerobic pathways with unlimited flux capacity.

Although the long-term, steady-state physiological responses to hypoxic stimulus were the same for both age groups, the stress of hypoxia-reoxygenation treatment showed short-term dysfunction in aging flies, across molecular to functional scales. Therefore, in addition to static differences in mitochondrial enzyme levels and activity, as others have measured, our results suggest that aging also affects the dynamic regulation of these enzyme fluxes in response to stress. One question that is opened by these data is whether the higher oxidative stress is creating new mitochondrial damage or merely making evident the effects of damage that has already accumulated with age. Also, are these results caused by the damage to the mitochondrial enzymes directly or to the ROS defenses that protect them?

Our approach compiled genome-scale data from several sources (the annotated fly genome, microarrays, and NMR metabolomics), along with detailed data from specific assays and the biochemical literature, and integrated them into a quantitative computer model that can be validated against future experiments. The model helped us to understand systems-level mechanisms for differences in the hypoxia response in young and old flies that both support and contribute to existing data regarding aging and mitochondrial dysfunction. In the future, specific molecular mechanisms can be further analyzed by comparing to similar models in other species and by perturbation analysis using selected enzyme mutations.

## Materials and Methods

### *Fly preparation*
Oregon-R wild-type flies were reared in constant light at 25° C and food was changed twice a week. Young flies (referred to as "3 day" or "3d" elsewhere in the text) were harvested for experiment at 3-5 days of age, and old flies ("40 day" or "40d" in the text) were collected at 38-42 days old. Because of the relatively small number of females surviving to 40 days of age, only males were used for experiment.

### *Hypoxia experiments*
The hypoxia experiments included five samples each of 3 conditions: control, 4-hour hypoxia, and 4 hours of hypoxia plus 5 minutes of recovery. A hypoxia chamber was created using four Sarstedt 50 mL plastic tubes. Two holes were drilled into the screw caps of the tubes and rubber hosing was inserted into each hole and sealed airtight with silicone adhesive. The hosing from the four tubes was then connected in parallel to a single inflow and a single outflow hose. At the time of experiment, filter paper was soaked in distilled water and placed in each tube to prevent drying. Approximately 50 flies were transferred into each tube, with two of the tubes containing young and two containing old flies. A mixture of nitrogen and 0.5% oxygen was then bubbled through distilled water and passed through the tubing into the vials. After circulating gas through the tubes for 15 minutes, the inflow and outflow hoses were sealed airtight with clamps, with the inflow sealed an instant before outflow in order to equalize the chamber pressure to the atmosphere. Control flies were similarly transferred from food vials and sealed in tubes with room air over the same time period in order to control for the effects of starvation and dehydration. Vials were lightly shaken to increase spacing among the immobile flies and stored on their side at 25º C. For the 5-minute recovery group, tubes were opened and exposed to room air after the 4-hour hypoxia duration.

For NMR and biochemical assays, vials were snap frozen in liquid nitrogen at the end of each time point and shaken to remove heads, legs, and wings. For each sample, 20 male thoraxes were separated from abdomen with microforceps on dry ice under a dissecting microscope and stored at -80° C until measurement.

### *Heart rate measurement*

Baseline heart rate was measured as described previously (Broderick et al., 2006; Paternostro et al., 2001). Briefly, 10 flies were anesthetized with triethylamine (Carolina Biological), and mounted on their backs on microscope slides using double-sided tape. Custom software and a motorized stage were used to locate and record the position of the heart, then draw a virtual line of pixels across the heart walls. Microscope recordings of the pixel values along this line were concatenated to create time-space (M-mode) image representations of heart wall motion, from which heart rate was extracted by custom image analysis algorithms.

The slide was placed in the custom hypoxia chamber for 4 hours as described above, then removed and M-mode images of the fly hearts were again recorded over the first 20 minutes of recovery. Since flies were exposed to room air simultaneously and therefore had to be measured in parallel, the software automatically multiplexed measurements by rotating through the 10 saved heart positions, recording four-second M-mode images each time. The time from exposure to room air was recorded alongside each image, then image data were binned into four 5-minute periods. Heart rate for each fly was calculated as the inverse of the average beat-to-beat interval over the first 5 minutes, normalized to baseline values, and this statistic was compared for the two age groups by the t-test.

### *Whole body recovery*

In whole-body recovery experiments, flies were exposed to hypoxia as described above, except that at the 4-hour time point flies were transferred to a lit surface where 10 males from each age group were chosen at random and separated with a paintbrush from the population. Recovery period, or the time from exposure to room air until the fly was standing upright on all legs, was recorded for each fly. Kaplan-Meier estimates and 95% confidence intervals for the cumulative distribution function were calculated using Matlab (MathWorks, Cambridge MA).

## NMR preparation

Thoraxes were homogenized in an ice bath for 3 minutes in 300μL of cold 1:1 acetonitrile:water buffer, using an OMNI TH homogenizer. Homogenates were centrifuged in a ice bath (4º C) for 10 minutes at 12,000 RPM. 10μL of the supernatant was used to determine the total protein concentration by the Bradford methods. For the Bradford assays, samples were diluted 10 times with extraction buffer. The supernatant was ultracentrifuged for 30 minutes at 8,500 RPM using Nanosep centrifugal devices (Pall Life Sciences, Ann Arbor, MI) with a 3 kDa molecular weight cutoff. To reduce the contamination by glycerol, a membrane wetting agent, to below 80μM, all Nanosep devices were washed 4 times (by 5 minutes centrifugation at 13,000 RPM) with 500μL deionized water. Filtrate was lyophilized using a vacuum centrifuge for 2 hours at 45° C. Samples were stored at -80° C until measured.

## NMR spectroscopy and data analysis

Dried samples were dissolved in 500μL $D_2O$ buffered at pH 7.4 with monobasic/dibasic sodium phosphate. The NMR standard TSP (3-trimethylsilyl-$^2H_4$-propionic acid) was added to the samples at a ratio of 1:100 by volume, resulting in a concentration of 0.488 mM. Analyses of samples were carried out by $^1H$ NMR spectroscopy on a Bruker Avance 500 operating at 500.13 MHz 1H resonance frequency. The NMR probe used was the 5 mm TXI 1H/2H-13C/15N Z GRD. All NMR spectra were recorded at 25° C. Typically $^1H$ were measured with 512 scans into 16384 data points, resulting in an acquisition time of 1.36 seconds. A relaxation delay of 2 seconds additionally ensured T1 relaxation between successive scans. Solvent suppression was achieved via the Noesypresat pulse sequence (Bruker Spectrospin Ltd.) in which the residual water peak is irradiated during the relaxation and mixing time of 80 μs. All $^1H$ spectra were manually corrected for phase and baseline distortions within XWINNMR$^{TM}$ (version 2.6, Bruker Spectrospin, Ltd.). Two-dimensional NMR methods including homonuclear correlation spectroscopy (TOCSY) and heteronuclear single quantum correlation spectroscopy (HSQC) were carried out in order to identify and subsequently confirm the assessment of metabolites. Peaks in the 1D spectra were identified, aligned, and quantified by "targeted profiling" algorithms (Weljie et al., 2006) within the software Chenomix NMR Suite 4.5 (Chenomix, Inc.). The list of metabolites discovered in the 2D spectra was used to guide quantification in one dimension.

## Standards and scaling factors for metabolite concentrations

In NMR spectra, absolute concentrations can be obtained from peak integrals if the sample contains an added internal standard of known concentration, or if the concentration of a substance is known by independent means (e.g., glucose determination by biochemical assay) (Beckonert et al., 2007). To determine absolute concentration of the 10 metabolites included in the model (alanine, lactate, glutamine, glutamate, glucose, pyruvate, proline, oxaloacetate and 4-aminobutyrate), a known concentration standard was acquired under the same experimental conditions and scaling factors were calculated for each metabolite. (see Table IV-S, Supplementary Material).

50μL of 10mM freshly made solution of each standard was added to 450uL of $D_2O$ buffered at pH 7.4 with monobasic/dibasic sodium phosphate containing 0.488mM of TSP (3-trimethylsilyl-$^2H_4$-propionic acid). Acquisition of the standards were carried out as described in the previous paragraph in duplicate, and quantified using the software Chenomix NMR suite 4.5 (Chenomix, Inc). The ratio $[Std]_{Chenomix}/[Std]_{solution}$ is defined as the scaling factor and is reported as the average the 2 experiments.

*ATP assay*

Twenty thoraxes from 3 days old flies or 40 days old males flies were homogenized in 300μL of 6M-guanidine-HCl in extraction buffer (100mM Tris-Acetate and 2mM EDTA, pH 7.75) to inhibit ATPase (Schwarze et al., 1998) and placed at 95°C for 5 minutes. The samples were then centrifuged in a cold room for 10 minutes at 12,000 RPM and the supernatant was diluted 500 times with the extraction buffer and mixed with luminescent solution (ATPLite, Perkin Elmer). The luminescence was measured by a luminometer (BT) and results were compared to the standards. The relative ATP level was calculating by dividing the luminescence by the total protein concentration.

*Glycogen and Trehalose assay*

Twenty thoraxes from 3 days old flies or 40 days old males flies were homogenized in an ice bath for 3 minutes in 300μL of 0.25M $Na_2CO_3$ using an OMNI TH homogenizer and incubated at 95°C for 2 hours to denature proteins. Aqueous solutions of 1M of acetic acid (150μL) and 0.2M of sodium acetate (600μL) were mixed with the homogenates and the suspensions were centrifuged for 10 minutes at 12,000 RPM. 100μl of the supernatant were placed in eppendorf to determine the glucose background. 200μL of supernatant were incubated overnight at 37°C with trehalase solution (0.05U/mL in 0.2M sodium acetate pH: 5.2).(Schulze et al., 1995) Glycogen was assayed using the method developed by Keppler and Decker (Keppler & Decker, 1974) with some modifications. 50μL aliquots were incubated with 500μL of an *A. niger* glucoamylase solution (8.7U/mL in 200mM of Acetate buffer pH: 4.8) for 2 hours at 40°C under constant agitation. The suspensions were centrifuged for 5 minute at 4000 RPM and glucose was determined on 20μL of supernatant by addition of 170μL of a G6-DPH (0.9U/mL)/ATP (1.6mM)/NADP (1.25mM) mixture in triethanolamine hydrochloride buffer (380mM TEA.HCl and 5.5mM of $MgSO_4$, pH: 7.5) and 10μL of Hexokinase solution (32.5U/μL in 3.2M ammonium sulphate buffer pH:6) and read at 340nm in a SpectraMax 190 (Molecular Device).

*Statistical analysis*

Student's t-test were performed to compare means between 2 samples, and *p* <0.05 were considered statistically significant. For analysis of the NMR data, Bonferroni t-tests were performed. Furthermore, for every metabolite with at least one measurement above 0.01 mM, an analysis of variance (ANOVA) was performed, with a Bonferroni correction applied to the *p*-value for the number of metabolites tested. Tukey's post-hoc tests were performed for metabolites, in which the null hypothesis of no change with treatment was rejected by ANOVA, for cross-comparison of each treatment. All statistical analysis were performed using GraphPad Prism software.

For the Principal Component Analysis, all metabolites with at least one measurement above 0.01 mM were included in the dataset. Each sample was normalized by protein content measured by the Bradford assay, and selected metabolites were scaled using standards as described above. Data from all samples (young and old; control, hypoxia and recovery) were combined into one matrix and principal components were computed using the princomp function in Matlab (Mathworks, Inc., Cambridge, MA). Principal component scores for the samples were plotted and visualized within Matlab.

### *Expanding the metabolic network reconstruction*

Our reconstruction of the central, ATP-generating metabolic network of Drosophila flight muscle, (described in (Feala et al., 2007), was expanded and refined using the absolute gene expression profile derived from an Affymetrix microarray of whole thorax in 3-day old flies (Girardot et al., 2006). Raw microarray data were combined with Affymetrix Drosophila Genome 2.0 annotation files to obtain gene identifiers, which were then linked to reactions and pathways of the KEGG database (Kanehisa et al., 2008; Kanehisa & Goto, 2000) using the dme_pathway.list and dme_enzyme.list batch files downloaded from ftp://ftp.genome.jp/pub/kegg/genes/organisms/dme. Genes from the microarray dataset were grouped by whether they had a KEGG identifier, and those existing in the KEGG database were further grouped by pathway. Mean expression levels in 3-day thorax were calculated for each KEGG pathway containing more than one Drosophila gene. Pathway expression levels were also visualized on KEGG pathway diagrams using the G-language Microarray System (Arakawa et al., 2005) (http://www.g-language.org/data/marray) on log-transformed expression data, which were re-scaled to range from 0 to 100 in order to fit the input format of the web service. The list of pathways with mean expression level greater 500 were visualized with this system and also investigated by a literature survey in order to determine whether to include the pathway in the model. The list of all Drosophila genes in KEGG was also sorted by thorax expression level and genes with expression levels greater than 500 were manually examined by literature and database search to determine inclusion in the model. Genes and reactions were entered into the model using the SimPheny biological database software (Genomatica, San Diego).

### *Flux-balance analysis*

Metabolite concentrations for the three experimental conditions (control, 4-hour hypoxia, 5-minute recovery) were converted into two sets of fluxes by dividing the differences in mean concentrations by the time period, resulting in units of nmol*mg prot$^{-1}$*min$^{-1}$. Standard errors (SE) of the metabolite fluxes were calculated from SE of the concentrations (using the formula $SE_{C2-C1} = \sqrt{[SE_{C1}^2 + SE_{C2}^2]}$ for subtracting random variables for concentration $C_1$ and $C_2$) and converted to the same units. Eleven compounds with measured hypoxia fluxes above .05 nmol*mg prot-1*min-1 were included in the model except for glycerol, which was contaminated by glycerol coating on the membrane filter, and β-alanine, a structural amino acid which saw a reverse flux during recovery that was unfeasible to incorporate in the current version of the model. Fluxes of glycogen and free glucose were similarly estimated from the biochemical assays. Metabolite pools were then simulated in the model by creating a sink for each compound and forcing fluxes into/out of the system to the values calculated from the data.

Flux-balance analysis was performed to simulate system flux distributions during hypoxia and recovery for both young and old flies. The objective function in all simulations was the reaction representing utilization of ATP via hydrolysis. The SimPheny software was used for initial flux-balance calculations and for visualizing superimposed fluxes on the metabolic network. All four simulations were exported to SBML, and are made available in the Supplementary Material as xml files.

We used Matlab (Mathworks, Inc., Cambridge MA) to analyze the sensitivity of flux distributions to variance in the data. The COBRA toolbox for constraint-based analysis (Becker et al., 2007) was used to import the SimPheny simulations and run flux-balance analysis within Matlab. Then, pseudo-random sets of fluxes were created by sampling normal distributions with mean and standard errors equal to those calculated for each metabolite flux. A group of 10,000 random flux sets was created for each of the four experimental conditions (old and young, recovery and hypoxia). Virtual "sinks" with unlimited capacity were created for each compound in order to represent metabolite pools, allowing intracellular accumulation and depletion in case substrates and end products did not perfectly balance. For each sampled set, fluxes into and out of the metabolite pools were constrained to the randomly selected fluxes and flux-balance analysis was performed.

## Acknowledgements

We would like to thank Matthew Owen for managing the fly stocks used for experiment, Rachel Nguyen for collecting survival data, and Francis Le for constructing the hypoxia chambers. This work was supported by the NIH grants BES-0506252 (McCulloch), P41-RR08605 (McCulloch), and R21-AG026729 (Paternostro).

**Figure Legends**

Figure 1: Recovery of physiological function in flies after severe hypoxic stress. (A) Whole body recovery was defined as time taken to recover from hypoxic stupor to a standing position. Old (40-day-old) flies took 2 to 3 times longer to recover than young (3-day-old) after 4 hours of 0.5% oxygen. N = 16 for both groups. (B) The hearts of young flies began beating again immediately on reoxygenation, while older fly hearts remained inactive for nearly 5 minutes. (inset) When heartbeats during the first 5 minutes were binned, 3-day-old heart rates were significantly higher (p<0.05). In both age groups, hearts had stopped by the end of the hypoxic period. N = 11 for both groups.

Figure 2: ATP concentrations in young and old fly thoraxes after a control period, at the end of hypoxia (4 hours at 0.5% $O_2$) and after 5 minutes of recovery. Concentrations are normalized to protein content. Differences are statistically significant (p<0.01) for control and recovery measurements, but not for hypoxia.

Figure 3: Glycogen and trehalose concentrations in young and old fly thoraxes after a control period, at the end of hypoxia (4 hours at 0.5% $O_2$) and after 5 minutes of recovery. Both concentrations are in terms of glucose monomer equivalents, normalized to soluble protein content. Differences between the ages are statistically significant (p<0.05) for recovery measurements, but not for hypoxia or control periods.

Figure 4: Principal component analysis of the metabolomics data. (A) Data samples plotted along the first two principal components (PCs) showed a large separation between young and old flies during recovery but not for the control or hypoxic conditions. The differences between young and old flies are most prominent on PC 1, while the different experimental oxygen conditions are best separated along PC 2. The plotted recovery points are for the 5 minute recovery data. (B) The principal components are a composite of metabolite concentrations. PC 1 is dominated by acetate concentrations, while PC 2 has highest contributions from lactate and alanine. Other metabolites contribute small amounts to the vector weighting. (inset) The first two principal components represent over 95% of the variation in the data.

Figure 5: Histograms of absolute thorax expression data, filtered through the KEGG pathway database. (A) Histogram of raw absolute expression data from 3-day-old thorax Affymetrix microarrays, after preprocessing and normalization described in (Girardot et. al., 2006) (B) Gene expression histogram for the subset of genes listed in at least one KEGG pathway. Note the accentuated "fat tail" of the filtered data, resulting in a slightly bimodal distribution. (C) Mean expression levels of thorax genes for the 122 KEGG pathways. Selected pathways are color-coded according to inclusion in the model (based

on literature data) and positioned above the bin that contains the corresponding average gene expression.

Figure 6: Fluxes calculated from NMR data and by flux-balance analysis. (Left) Fluxes during 4-hour hypoxia period. (Right) Fluxes during 5-minute post-hypoxic recovery. (Top) Fluxes calculated from NMR concentration profiles. These fluxes were applied to the model as constraints. Error bars mark standard errors derived from concentration measurements. (Bottom) Fluxes calculated in flux-balance simulations. Error bars mark standard errors of simulations in sensitivity analysis. Compound abbreviations: glcgn: glycogen, ac: acetate, ala: alanine, lac: lactate, abut: 4-aminobutyrate, glc: free glucose, glu: glutamate, gln: glutamine, oaa: oxaloacetate, pro: proline, pyr: pyruvate, o2: oxygen, co2: carbon dioxide, h2o: water, h: protons, nh4: ammonia. Reaction abbreviations: atps: ATP synthase, atp: total ATP production, nadh: NADH dehydrogenase, sucd: succinate dehydrogenase (Complex II), cyoo: cytochrome oxidase, pfk: phosphofructokinase, gpdh: glycerol-3-phosphate dehydrogenase, pyk: pyruvate kinase, pdh: pyruvate dehydrogenase, cs: citrate synthase, sucd: succinate dehydrogenase (TCA cycle), mdh: malate dehydrogenase.

Figure 7: Flux map comparing recovery fluxes in young versus old flies. Fluxes of young and old are represented by the red-green color scale, while color coding from black to yellow indicates a ratio of 1 over a large range of absolute flux values. Numerical flux values are printed next to reaction abbreviations. Major anaerobic end products are shown in blue.

**Table 1: Comparison of Drosophila models**

|  | **Version 0.5** <br> **(Feala *et. al.*, 2007)** | **Version 1** <br> **(current)** |
|---|---|---|
| Genes | 169 | 211 |
| Reactions | 171 | 196 |
| Metabolites | 76 | 83 |
| Pathways | Glycolysis <br> TCA cycle <br> Oxidative phosphorylation <br> Fatty acid oxidation <br> Proline degradation <br> Alanine/glutamate met <br> ROS detoxification | Glycolysis <br> Gluconeogenesis <br> Pentose phosphate shunt <br> TCA cycle <br> Oxidative phosphorylation <br> Fatty acid oxidation <br> Proline degradation <br> Alanine/glutamate metabolism <br> Glutamine metabolism <br> Tyrosine/phenylalanine metabolism <br> Aminosugar metabolism <br> ROS detoxification <br> Starch and sucrose metabolism |
| Average confidence <br> 0-4 scale <br> (non-transport rxns) | 1.83 | 2.4 |

**Table 2: Averages of key fluxes in the simulations**

|  | 3d hypoxia | 40d hypoxia |  | 3d recovery | 40d recovery |
|---|---|---|---|---|---|
| **Glucose** | - 3.6 | - 3.6 |  | - 19.4 | - 53.3 |
| **Oxygen** | - 11.5 | - 11.8 |  | - 576 | - 492 |
| **ATP** | 64.5 | 67.6 |  | $3.17*10^3$ | $2.89*10^3$ |
| **H+** | 4.1 | 4.1 |  | -135 | 55.3 |

(-) = consumption; (+) = production.   glucose = free glucose + glycogen <br>
units: nmol/min/mg prot

## Post-hypoxia recovery (after 4 hours @ 0.5% O$_2$)

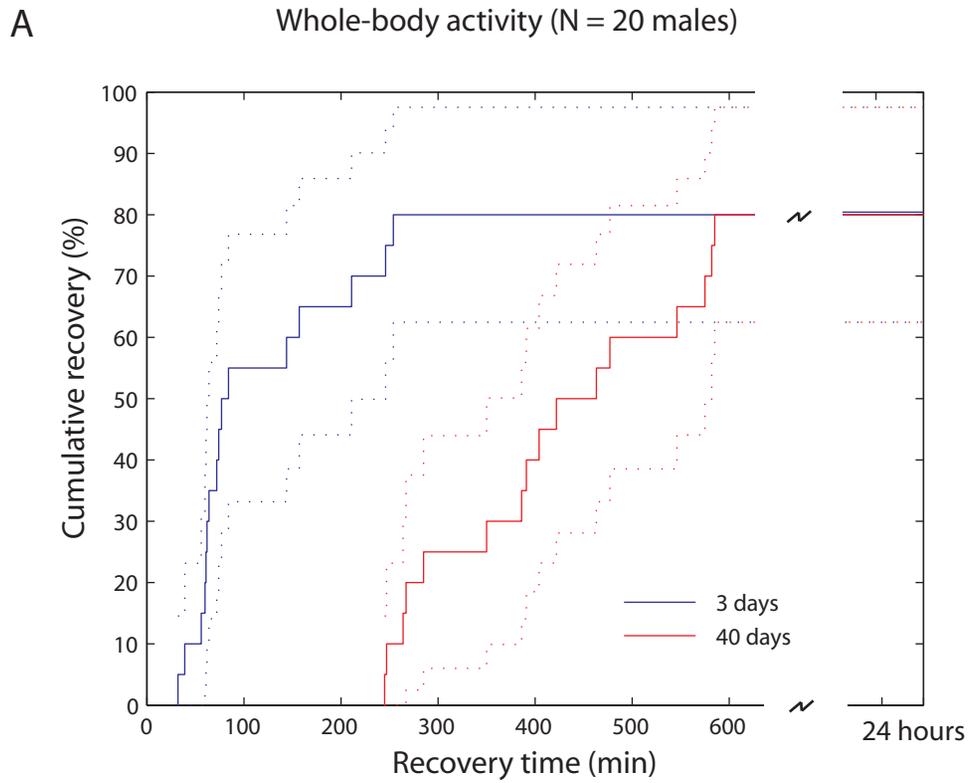

A. Whole-body activity (N = 20 males)

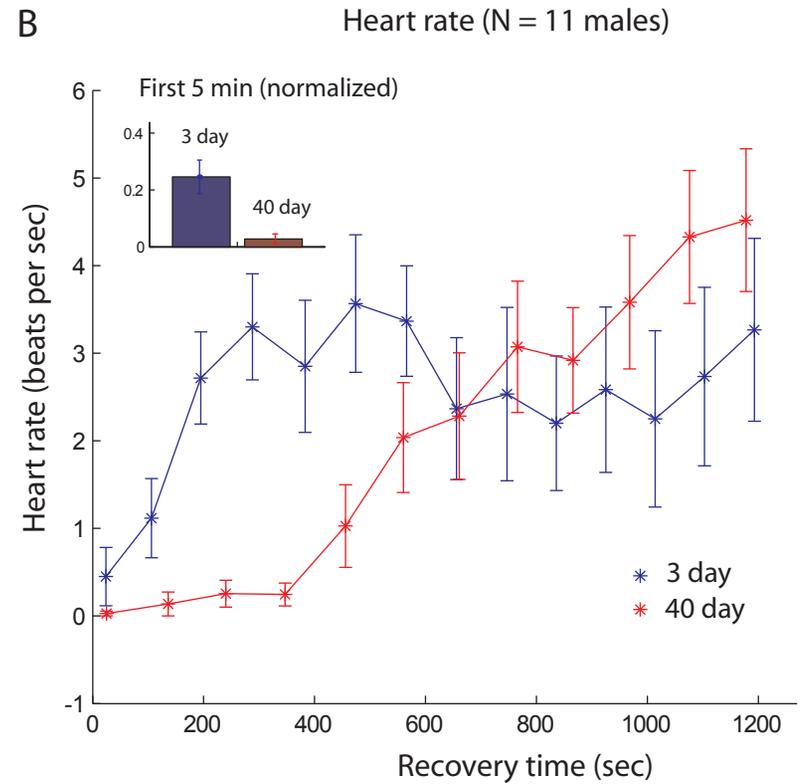

B. Heart rate (N = 11 males)

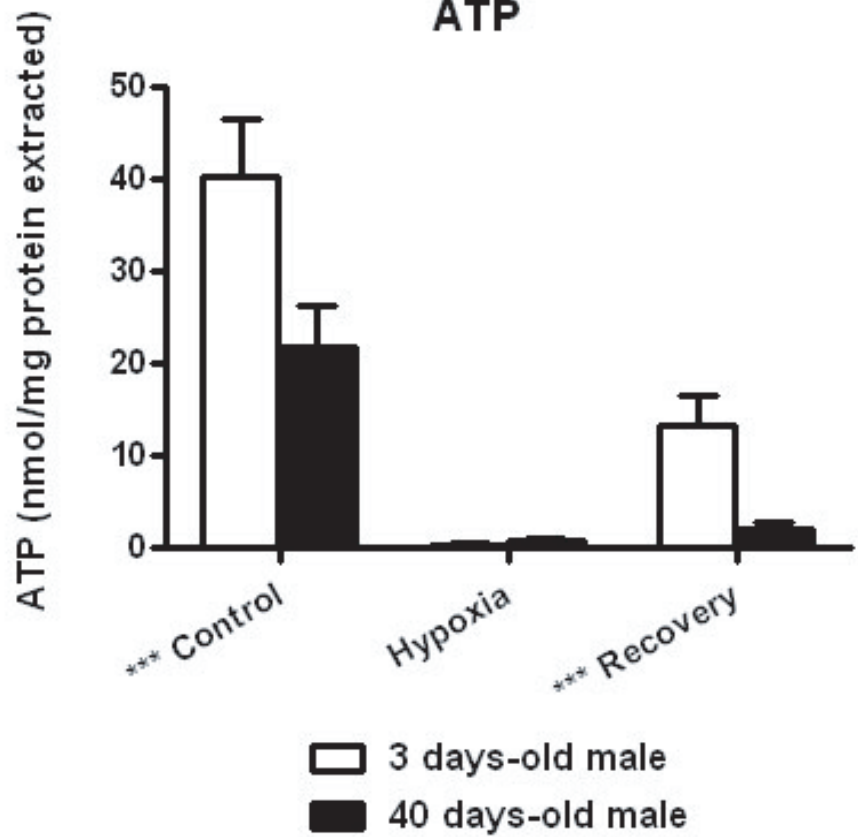

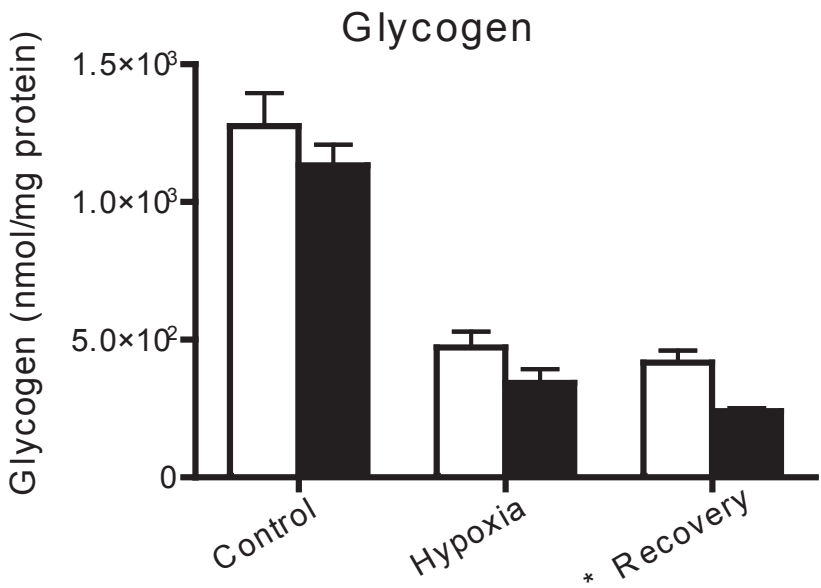

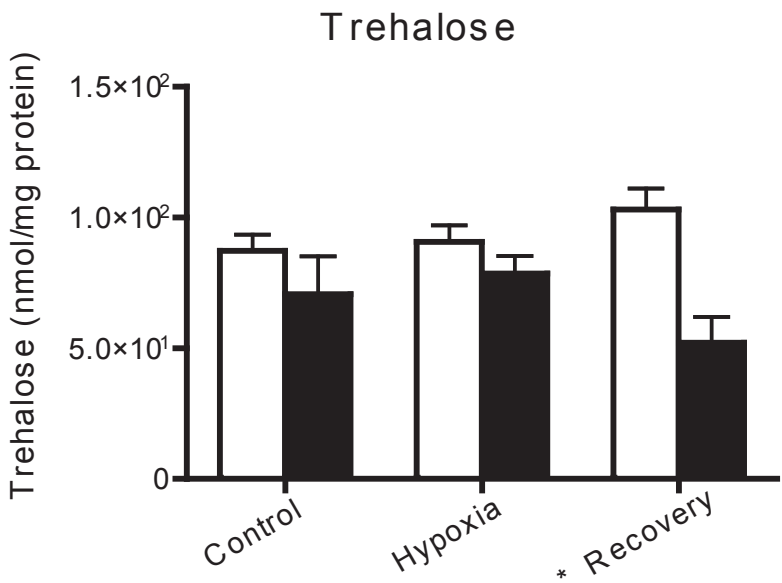

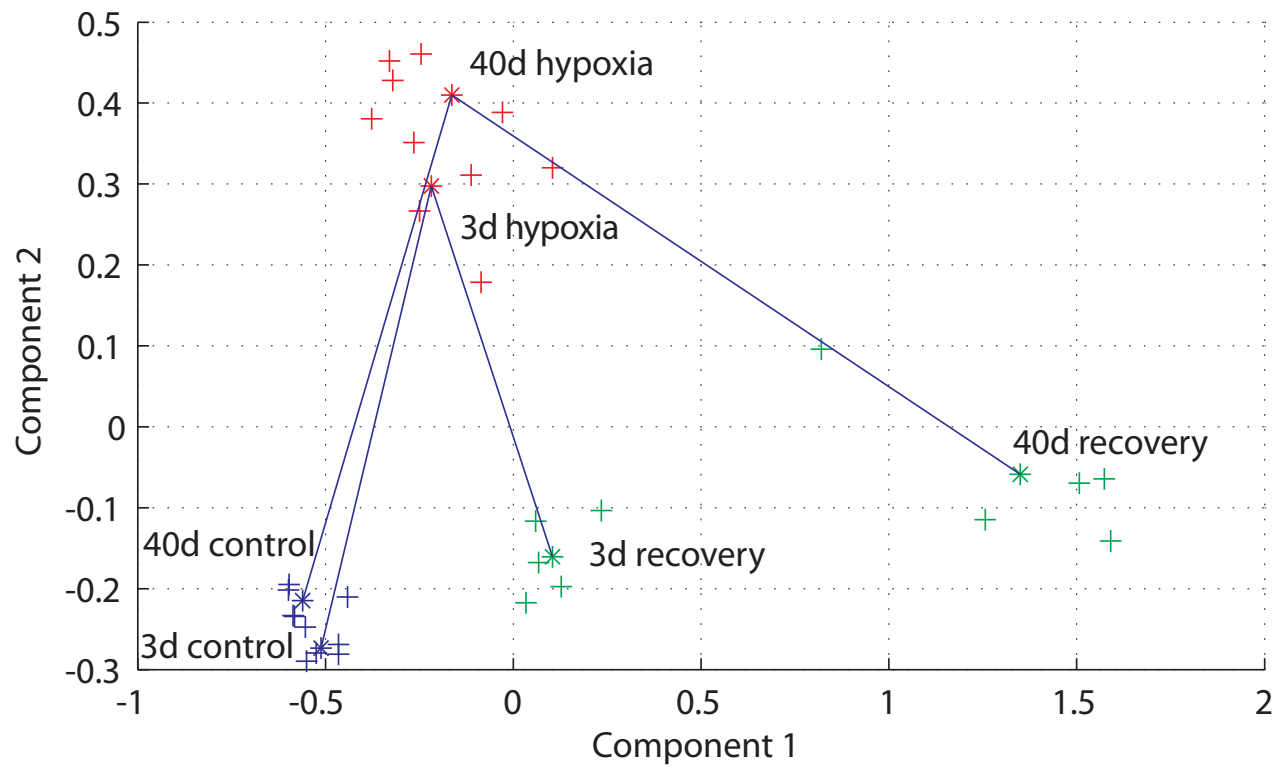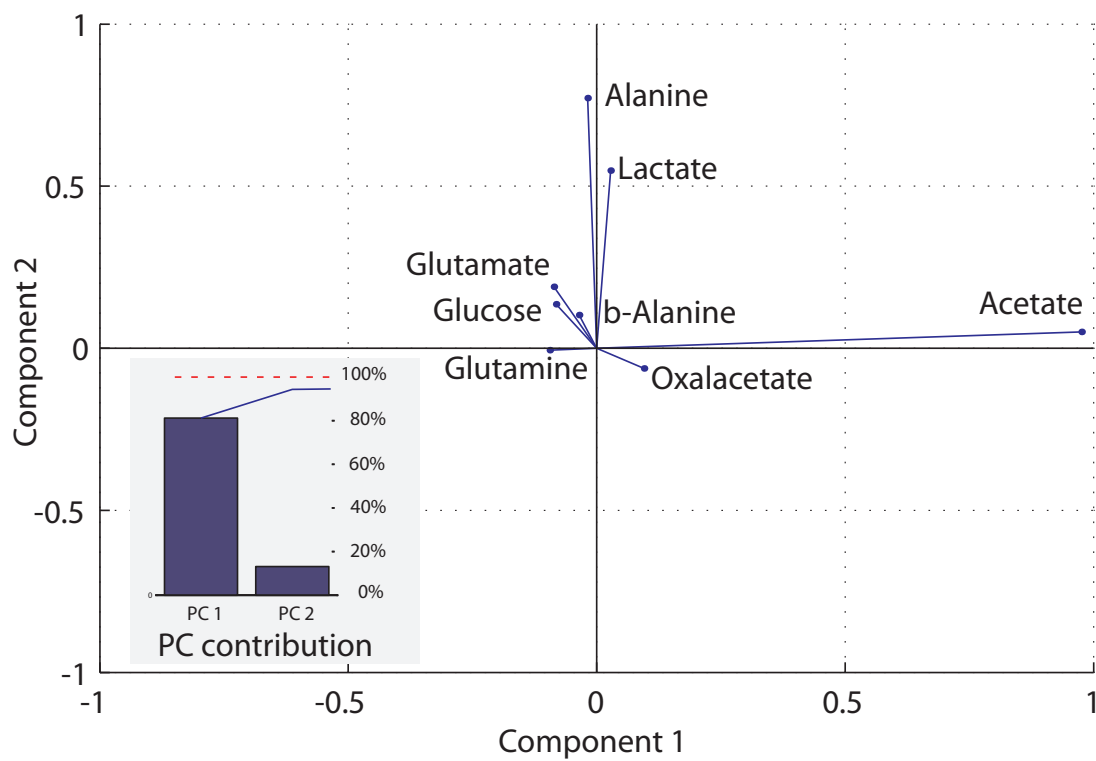

# Thorax gene expression (3-day old adult)

**A** All genes

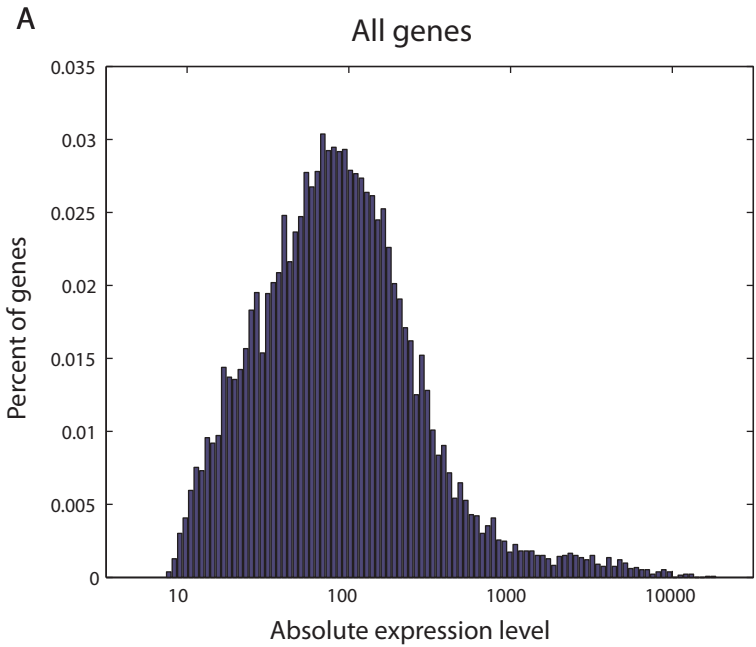

**B** Genes listed in KEGG pathways

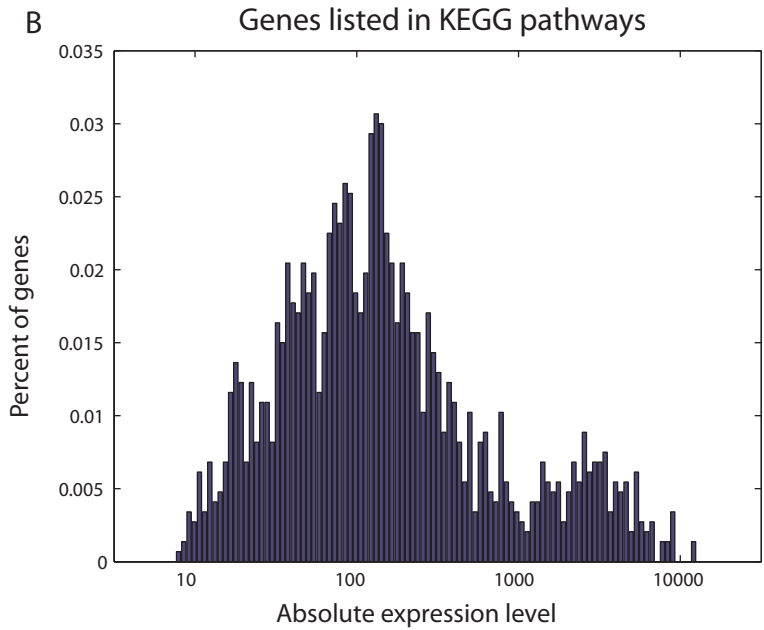

**C** Mean expression of KEGG pathways

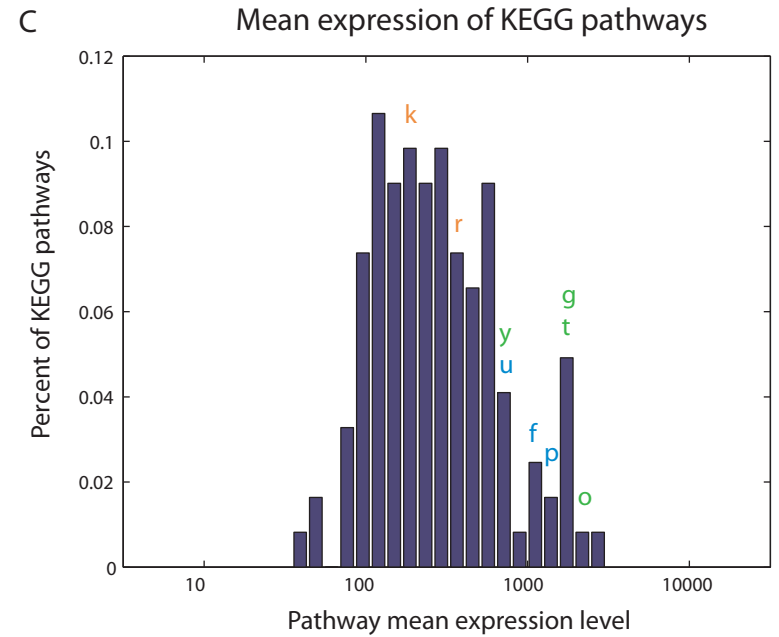

Graph codes (above):

Pathways already in model:
  y: Pyruvate metabolism pathway
  g: Glycolysis pathway
  t: Tricarboxylic acid cycle
  o: Oxidative phosphorylation pathway

Pathways to be added:
  p: Pentose phosphate pathway
  f: Fructose and mannose metabolism
  u: Glutamate and glutamine metabolism

Pathways excluded from model:
  r: Urea cycle
  k: Degradation of ketone bodies

# Model flux summary

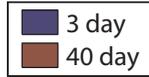

## Hypoxia (4 hrs)

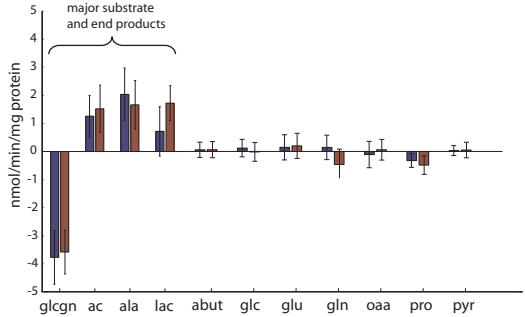

## Recovery (5 mins)

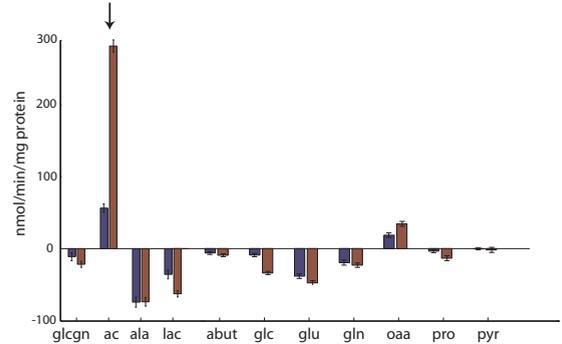

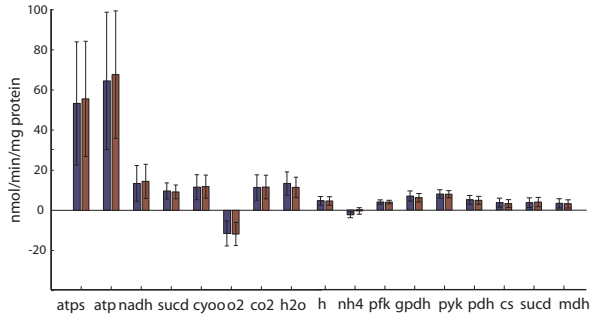

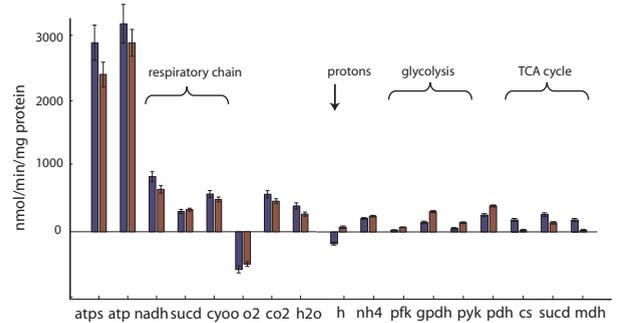

# Flux balance analysis of recovery from hypoxia: Young versus old flies

(simulations based on NMR fluxes at 5 minutes reoxygenation after 4 hours at 0.5% oxygen)

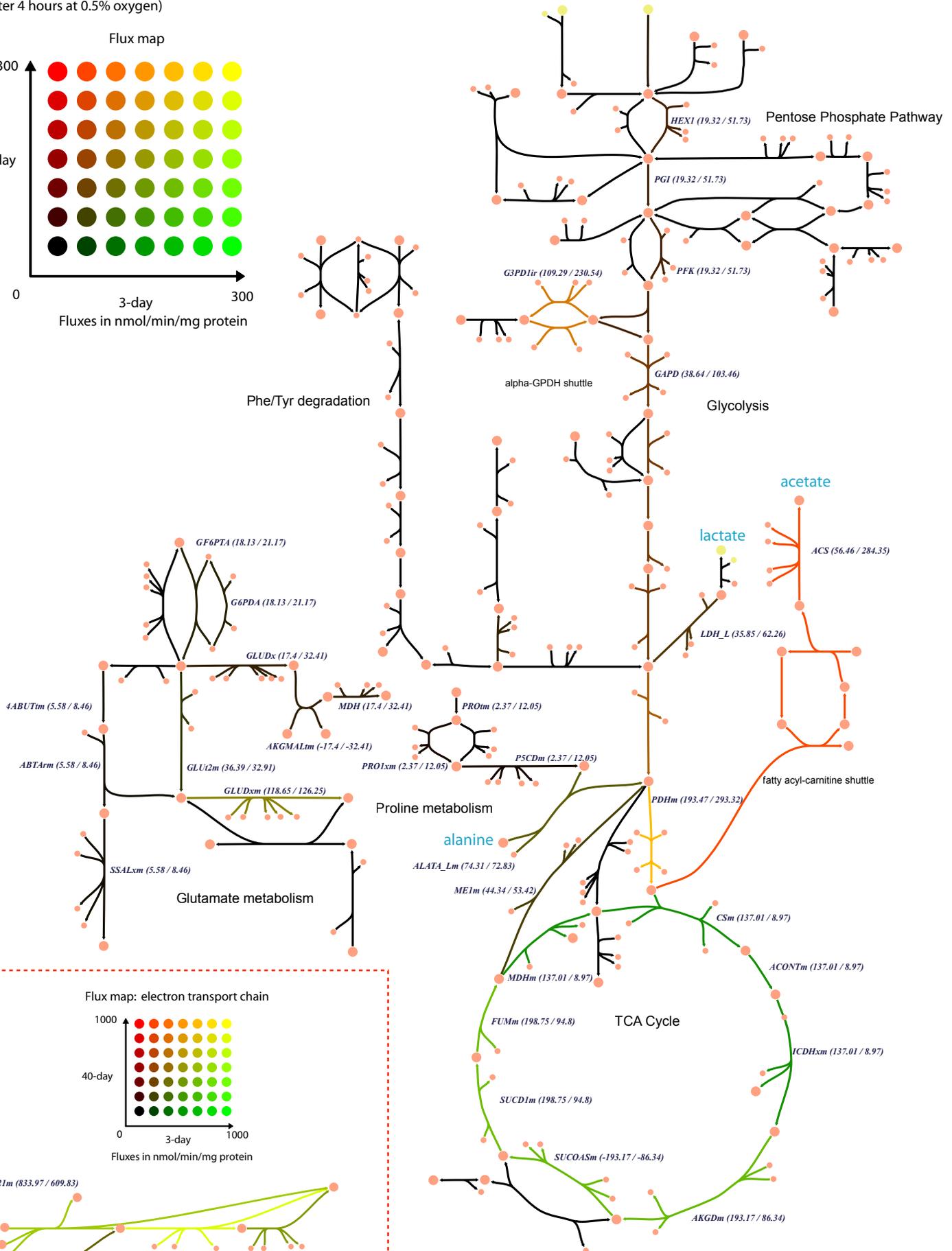